\documentclass[
    ,final            
  ]
  {aipproc}

\layoutstyle{6x9}

\newcommand{\charged}{\pi^+\pi^-\pi^-}
\newcommand{\neutral}{\pi^-\pi^0\pi^0}
\newcommand{\atwo}{2^{++}1^+(\rho)D}
\newcommand{\aone}{1^{++}0^+(\rho)S}
\newcommand{\pitwo}{2^{-+}0^+(f_2)S}
\newcommand{\pione}{1^{-+}1^+(\rho)P}
\newcommand{\pitwofa}{2^{-+}0^+(\rho)F}
\newcommand{\pitwofb}{2^{-+}1^+(\rho)F}
\newcommand{\pitwos}{2^{-+}1^+(\sigma)D}
\newcommand{\aonehigh}{1^{++}1^+(f_2)P}

\newcommand{\mevcc}{\!\mathrm{MeV}/c^2}

\newcommand{\gevc}{\!\mathrm{GeV}/c}

\newcommand{\gevt}{\!\mathrm{GeV}^{2}/c^2}

\newcommand{\pim}{\pi^-}

\newcommand{\jpc}{J^{PC}}

\newcommand{\onemp}{1^{-+}}

\newcommand{\twomp}{2^{-+}}

\newcommand{\twopp}{2^{++}}

\begin{document}

\title{PWA of $3\pi$ Final States and a \\
Search for the $\pi_1(1600)$}



\author{Ryan Mitchell}{
  address={Department of Physics, Indiana University, Bloomington, IN 47405}
}

\begin{abstract}
Partial wave analyses (PWA) of the $3\pi$ systems in the reactions $\pim p \rightarrow \charged p$ (the ``charged'' mode) and $\pim p \rightarrow \neutral p$ (the ``neutral'' mode) with an 18.3~$\gevc$ pion beam were performed using high statistics data from the E852 experiment. Conventional signals, such as the $a_1(1260)$, the $a_2(1320)$, and the $\pi_2(1670)$, were found to be remarkably stable to the choice of waves included in the fit.  In contrast, possible evidence for the $\pi_1(1600)$ in the $\jpc = \onemp$ exotic wave amplitude disappears when additional decay modes of conventional mesons (especially those of the $\pi_2(1670)$) are included in the PWA fit. \end{abstract}

\maketitle


\section{Introduction}

A $q\bar{q}$ bound state is constrained to a series of allowed quantum numbers $\jpc$, where $J$ is the spin of the bound state, $P$ its parity, and $C$ its symmetry under charge conjugation.  If, however, the gluonic field between the quark and antiquark is excited (forming a ``hybrid'' meson) then $\jpc$ combinations are allowed that would have otherwise been forbidden.  Finding mesons with these ``exotic'' quantum numbers provides experimental input to our theoretical understanding of quark confinement.

One candidate for a $\jpc = \onemp$ exotic meson is the $\pi_1(1600)$ decaying to $\rho\pi$ in a $P$-Wave.  This state was first reported by the E852 Collaboration in 1998~\cite{Adams98,Chung02} in an analysis based on 250K $\pim p \rightarrow \charged p$ events collected during the 1994 run of the E852 experiment.  The analysis presented here uses 2.6M events (after selection cuts) of the type $\pim p \rightarrow \charged p$ (``charged'' mode) and 3.0M $\pim p \rightarrow \neutral p$ (``neutral'' mode) events collected during the 1995 E852 run.



\section{The E852 $3\pi$ Data Sets}

The E852 experiment ran at the Alternating Gradient Synchrotron (AGS) at the Brookhaven National Lab during the years 1994-1995.  An 18.3~$\gevc$ pion beam was incident on a liquid hydrogen target.  Charged particles were tracked downstream of the target with a series of proportional wire chambers (PWC) and drift chambers.  The positions of recoiling protons were detected with a central cylindrical drift chamber (TCYL).  Photons were measured with a downstream lead-glass calorimeter (LGD).  Details can be found elsewhere~\cite{Teige99}.

\begin{figure}
  \includegraphics[width=.55\textwidth]{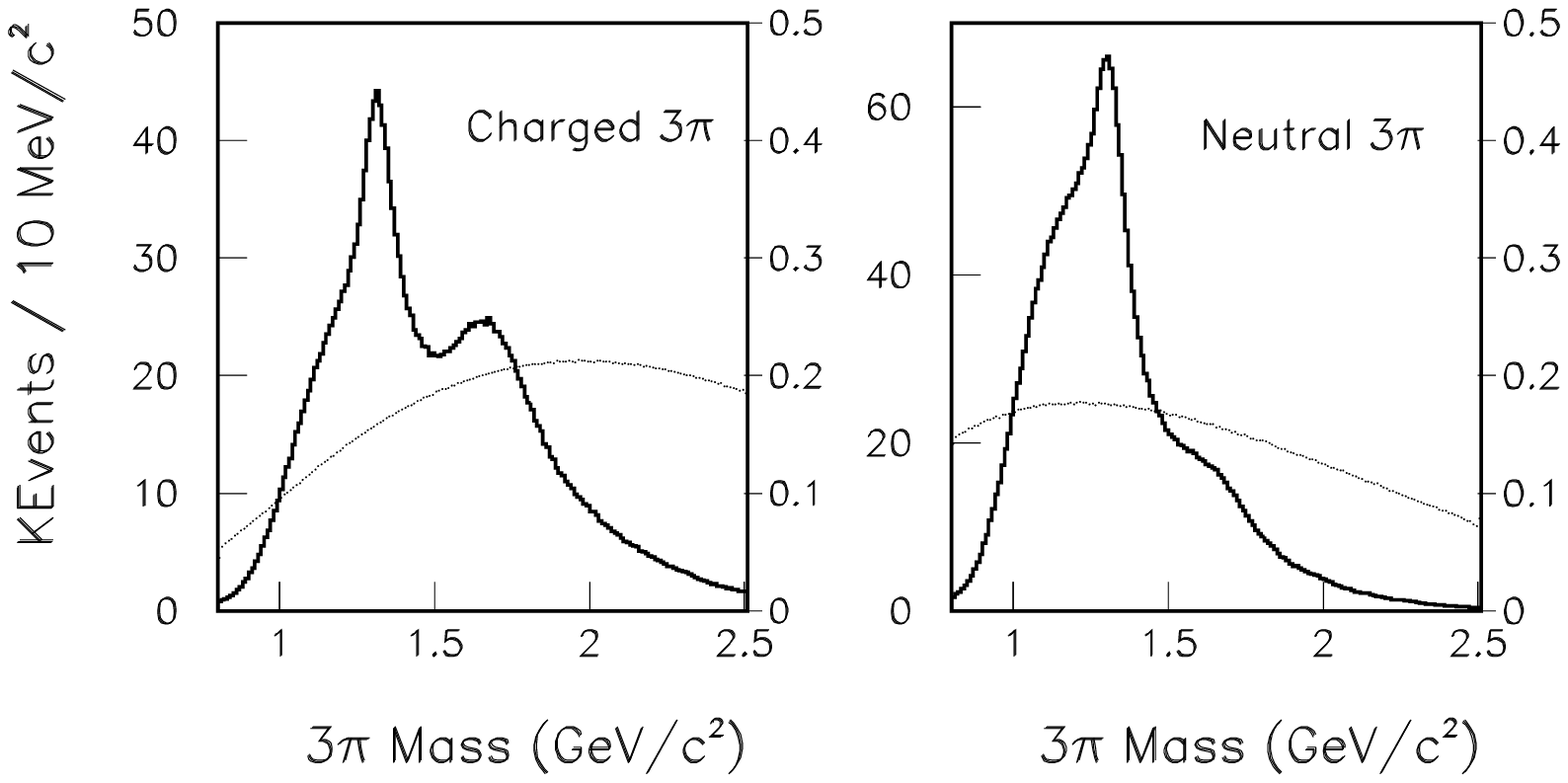}
  \includegraphics[width=.55\textwidth]{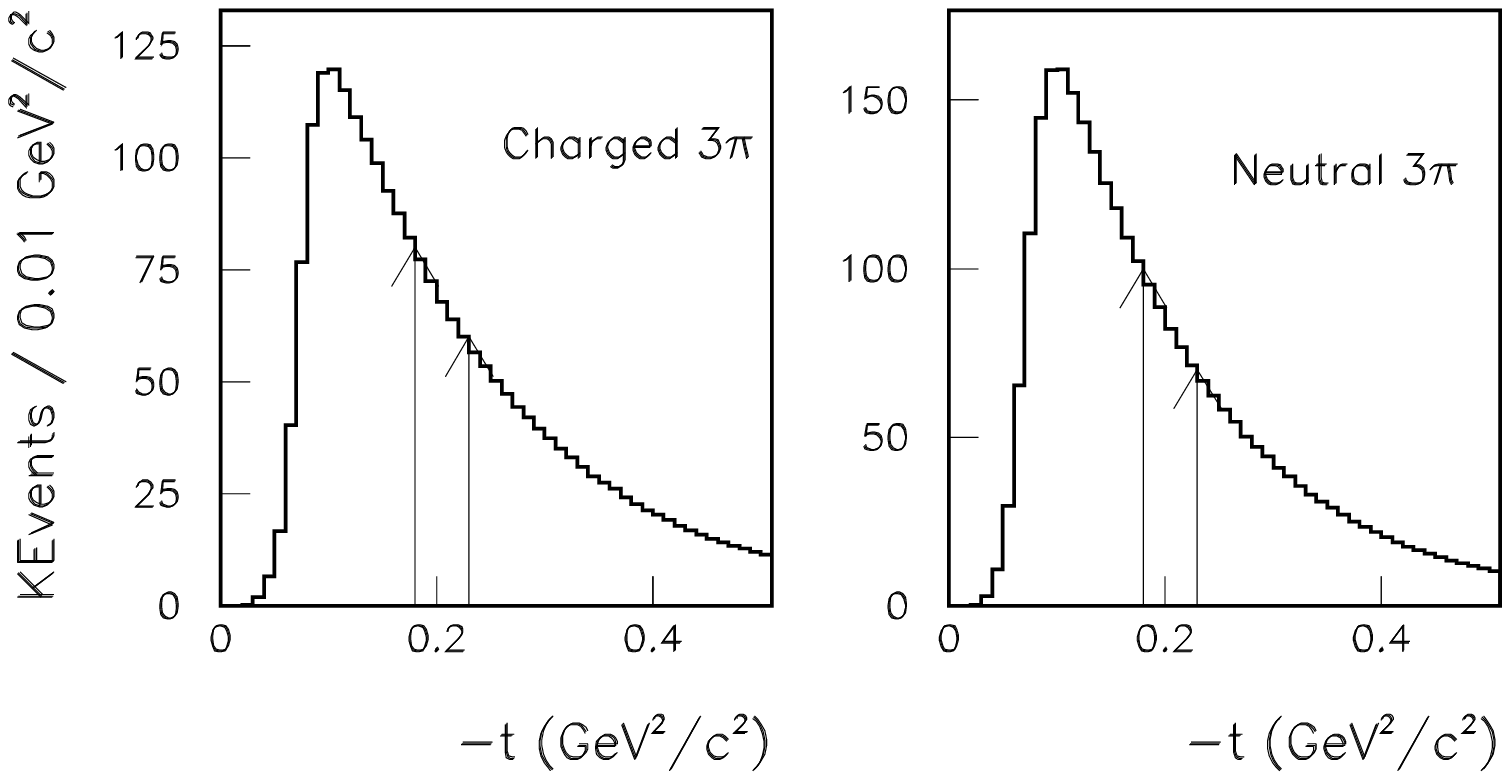}
  \caption{Uncorrected mass and $-t$ distributions for the charged and neutral modes.  Mass acceptances are superimposed on the mass plots (the scale is given on the right axes).}
  \label{raw}
\end{figure}

The trigger for the charged $3\pi$ mode required three tracks in the downstream PWCs in coincidence with one track (the recoil) in the central TCYL.  The neutral mode trigger\footnote{A global inefficiency of the triggering PWCs resulted in an overall normalization factor between the two charge modes.  This overall inefficiency has no effect within a given trigger and was not modeled in the Monte Carlo.  To bring the results from the charged and neutral mode triggers into line, all results in the charged mode were scaled by a factor of 1.3.} required one track in the PWCs, one track in the TCYL, and energy in the LGD inconsistent with a single $\pi^0$.  Further selection cuts included requiring a single vertex in the target, a confidence level requirement of a kinematic fit, and consistency between the angle of the missing recoil momentum and the position of the TCYL track~\cite{3piE852}.

Figure~\ref{raw} shows the uncorrected $3\pi$ mass distributions for the two charge modes.  Structures corresponding to the $a_1(1260)$, $a_2(1320)$, and $\pi_2(1670)$ are evident. The momentum transfers ($-t$) between the incoming pion and the $3\pi$ systems are also shown in Figure~\ref{raw}. The data was divided into 12 different bins of $-t$.  Results shown in this paper are from $0.18$ to $0.23~\gevt$, the region in Figure~\ref{raw} marked by arrows, and the region where there is maximum sensitivity to the largest number of $3\pi$ states.

\section{Partial Wave Fit Procedure}

The formalism used in fitting the data was identical to that used in the earlier E852 analysis~\cite{Chung02}.  The $3\pi$ intensity was expanded in each mass bin in a series of partial waves, each of which, in the isobar model, describes a resonance decay to an isobar and a bachelor $\pi$ and the subsequent isobar decay to $\pi\pi$.  Individual partial waves are denoted by $J^{PC}M^{\varepsilon}(isobar)L$, where $\jpc$ are the quantum numbers of the resonance, $M$ is its spin projection, $L$ is the orbital angular momentum between the isobar and bachelor $\pi$, and $\varepsilon$ is the reflectivity (corresponding to the naturality of the exchange particle in the production mechanism). 

\begin{table}
\caption{\label{thewaves}Partial waves under consideration in this analysis. }
\begin{tabular}{|lc|lc|lc|}
\hline
$J^{PC}M^{\varepsilon}(isobar)L$ & Wave Set\footnote{Indicates whether wave was used in the high wave
set [H], the low wave set [L] or in both [B].}
 & $J^{PC}M^{\varepsilon}(isobar)L$ & Wave Set & $J^{PC}M^{\varepsilon}(isobar)L$ & Wave Set  \\
\hline
$  0^{-+}0^+(\sigma)S$ & B &
  $  2^{++}1^+(\rho)D$ & B &
    $  2^{-+}1^+(f_2)D$  & B\\
$  0^{-+}0^+(f_0)S$ & B &
  $  2^{++}0^-(\rho)D$ & B&
    $  2^{-+}0^+(f_2)D$  & B \\
$  0^{-+}0^+(\rho)P$ & B&
  $  2^{-+}0^+(f_2)S$  & B &
    $  2^{-+}0^+(\rho)F$ & H \\
$  1^{++}0^+(\rho)S$  & B &
  $  2^{-+}1^+(f_2)S$  & B &
    $  2^{-+}1^+(\rho)F$ & H\\
$  1^{++}1^+(\rho)S$  & B &
  $  2^{-+}1^-(f_2)S$  & L&
    $  3^{++}0^+(\rho_3)S$ & B \\
$  1^{++}1^-(\rho)S$   & L&
  $  2^{-+}0^+(\rho)P$  & B &
    $  3^{++}0^+(f_2)P$  & H \\
$  1^{++}0^+(\sigma)P$  & H &
  $  2^{-+}1^+(\rho)P$ & H &
    $  3^{++}0^+(\rho)D$  & H\\
$  1^{++}0^+(f_0)P$  & H &
  $  2^{-+}0^+(\rho_3)P$  & H&
    $  4^{++}0^+(\rho_3)D$ & H \\
$  1^{++}1^+(f_2)P$  & H&
  $  2^{-+}1^+(\rho_3)P$  & H &
    $  4^{++}0^+(f_2)F$  & H \\
$  1^{++}0^+(\rho)D$ & L &
  $  2^{-+}0^+(\sigma)D$  & B &
    $  4^{++}0^+(\rho)G$  & H\\
$  1^{-+}1^+(\rho)P$  & B &
  $  2^{-+}0^+(f_0)D$  & H&
    $  4^{-+}0^+(\rho_3)P$  & H \\
$  1^{-+}0^-(\rho)P$  & B&
  $  2^{-+}1^+(\sigma)D$  & H &
    $  4^{-+}0^+(\rho)F$  & H \\
$  1^{-+}1^-(\rho)P$  & B &
  $  2^{-+}1^+(f_0)D$ & H &
    Background &  B\\
\hline
\end{tabular}
\end{table}

A set of waves to include in the fit was chosen by the following procedure.  First, a parent wave set was defined that included waves with the $\rho(770)$, $\rho_3(1690)$, $f_0(980)$, $f_2(1270)$ and $\sigma$ isobars, where $\sigma$ denotes the $\pi\pi$ $S$-Wave amplitude defined in~\cite{Chung02}.  Waves were restricted to those with $J \le 4$ and $M \le 1$.  Waves were then removed one by one from the fit and the resulting changes in log(Likelihood) were examined.  If the change was greater than 40 units, the wave was judged significant and was returned to the wave set\footnote{Exceptions to this method were the $\onemp$ waves, which were judged to be insignificant by the likelihood criteria, but were kept for comparison with the earlier results.}.  A set of 35 waves survived this procedure.  See~\cite{3piE852} for details.

A comparison of the resulting wave set (termed the ``high wave set'') and the wave set used in the original E852 $3\pi$ analysis (the ``low wave set'') is given in table~\ref{thewaves}.  The low wave set included 20 waves.

\section{Results}

\begin{figure}
  \includegraphics[width=.55\textwidth]{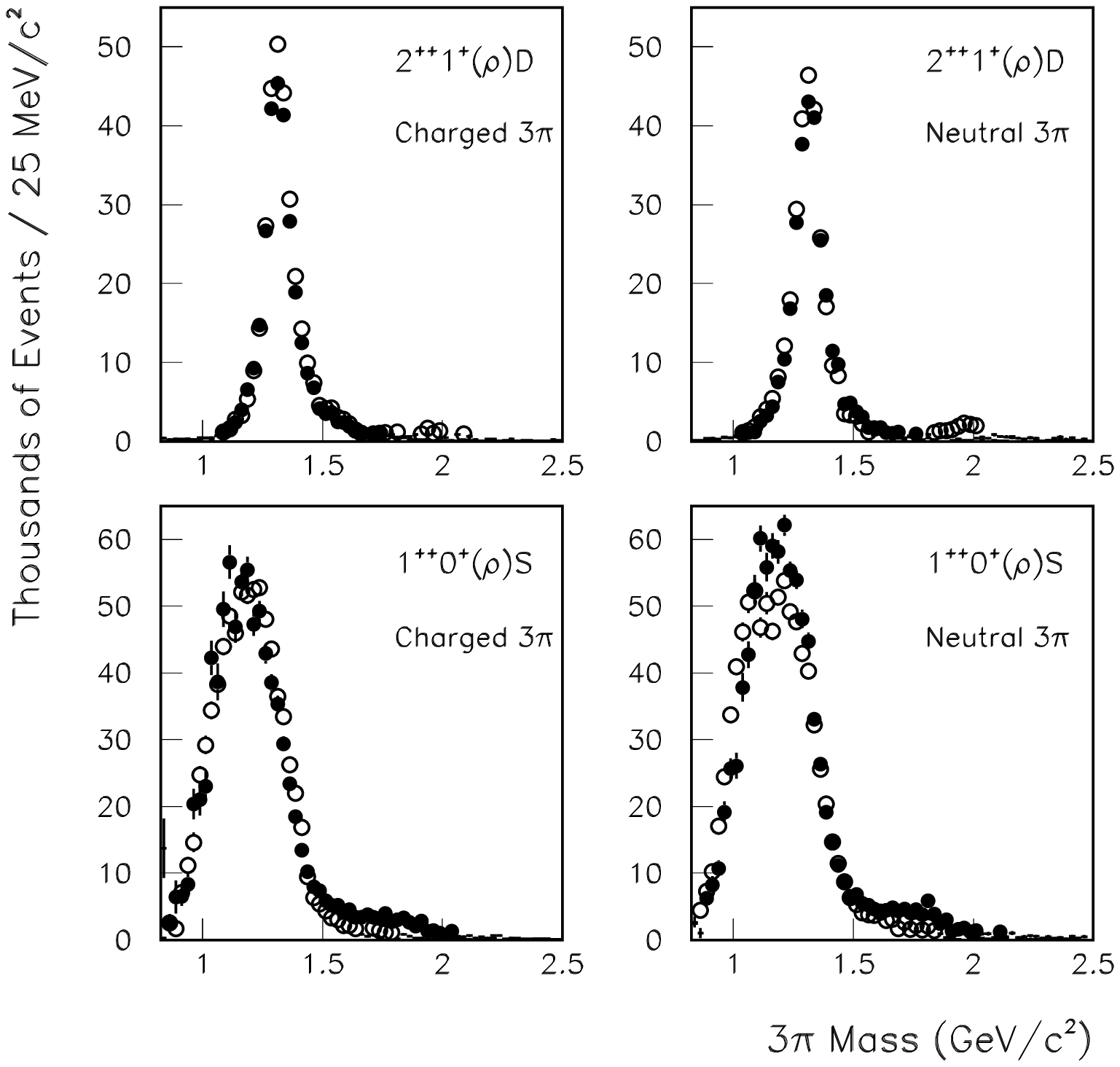}
  \includegraphics[width=.55\textwidth]{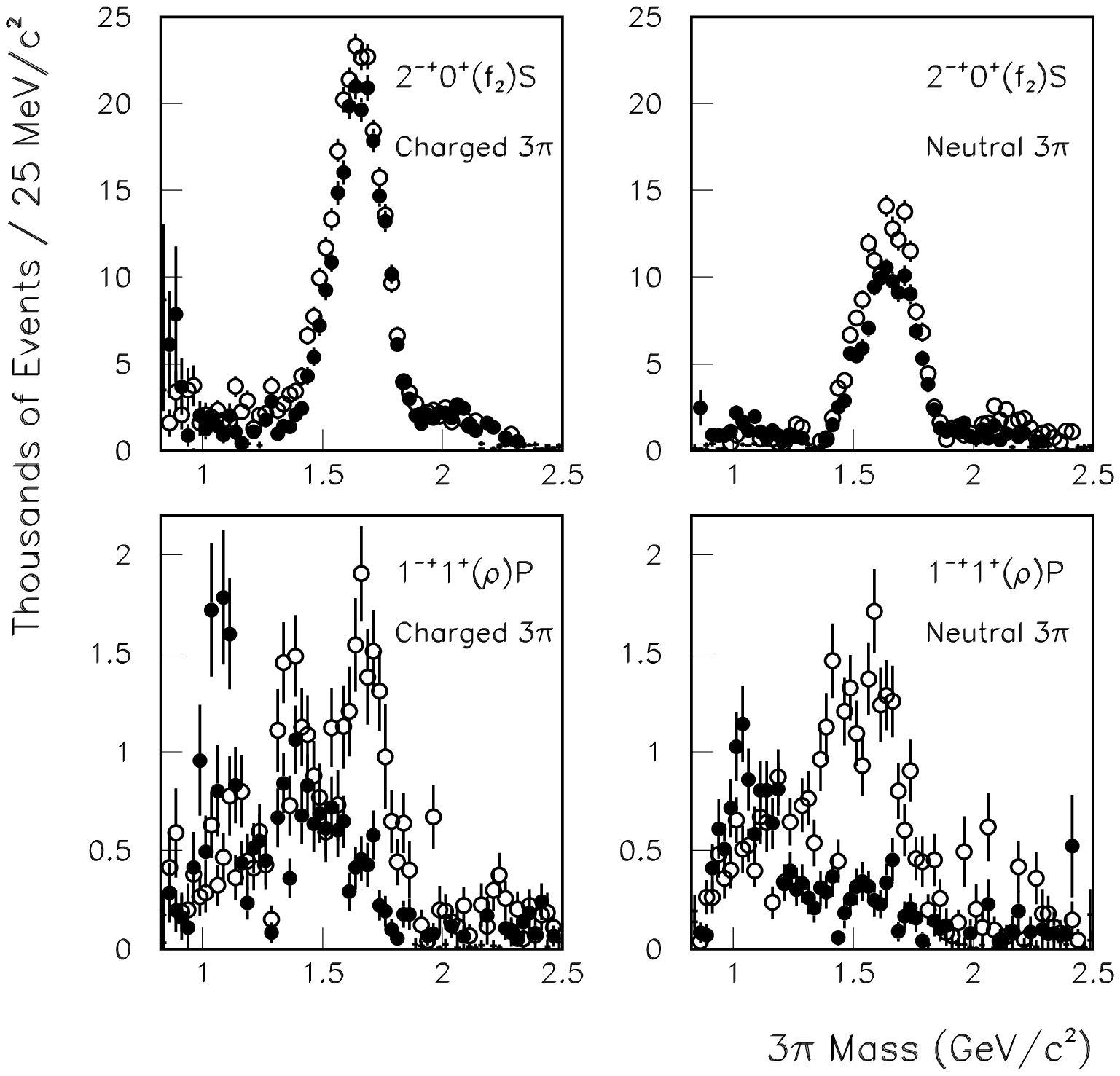}
  \label{intensities}
  \caption{Intensities of the $\atwo$, the $\aone$, the $\pitwo$, and the $\pione$ waves resulting from fits with the high (solid circles) and low (open circles) wave sets.}
\end{figure}

\begin{figure}
  \includegraphics[width=.55\textwidth]{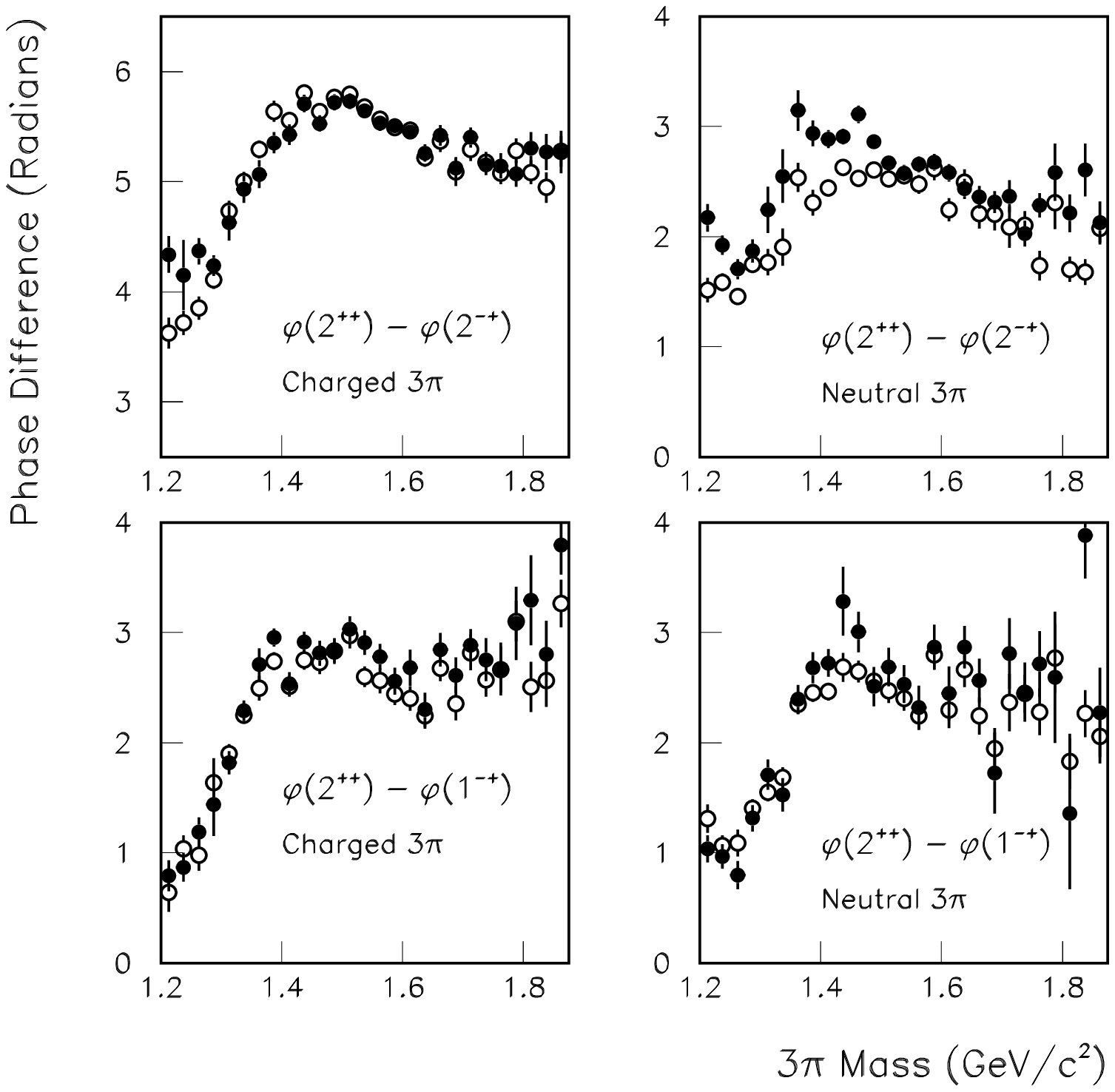}
  \includegraphics[width=.55\textwidth]{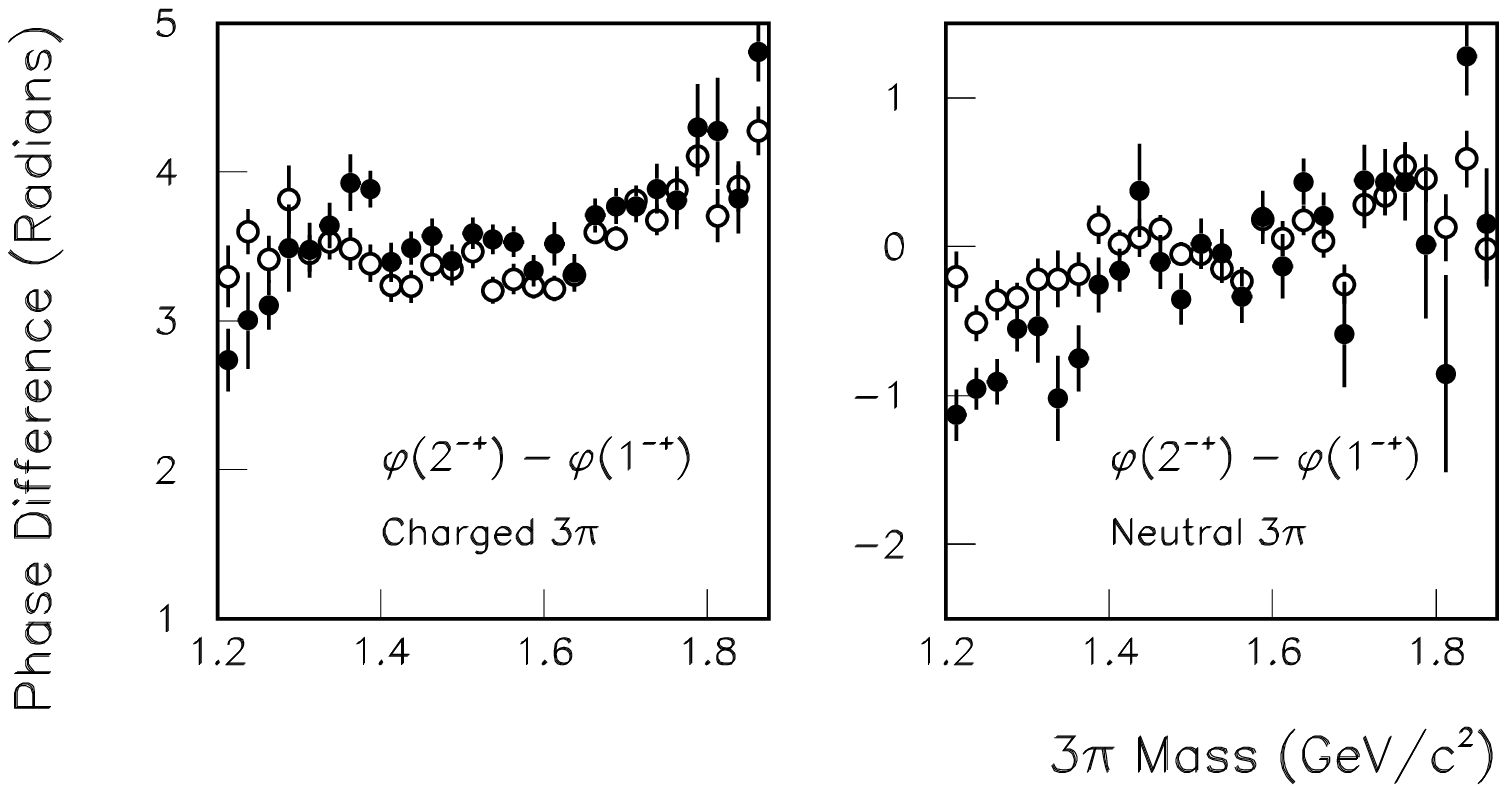}
  \caption{Phase differences between the $\pione$, the $\atwo$, and the $\pitwo$ waves resulting from fits with the high (solid circles) and low (open circles) wave sets.}
  \label{phases}
\end{figure}

\begin{figure}
  \includegraphics[width=.55\textwidth]{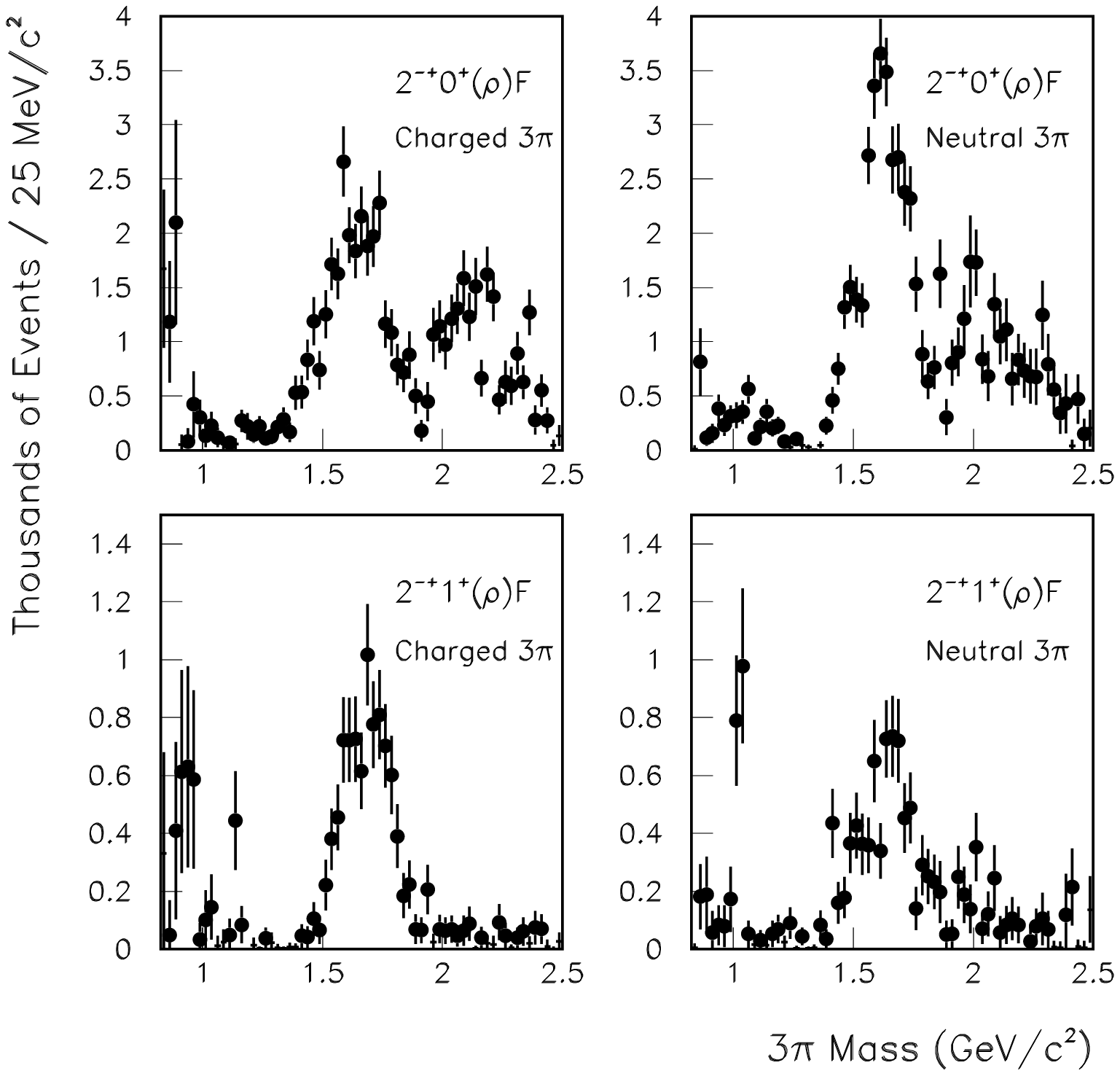}
  \includegraphics[width=.55\textwidth]{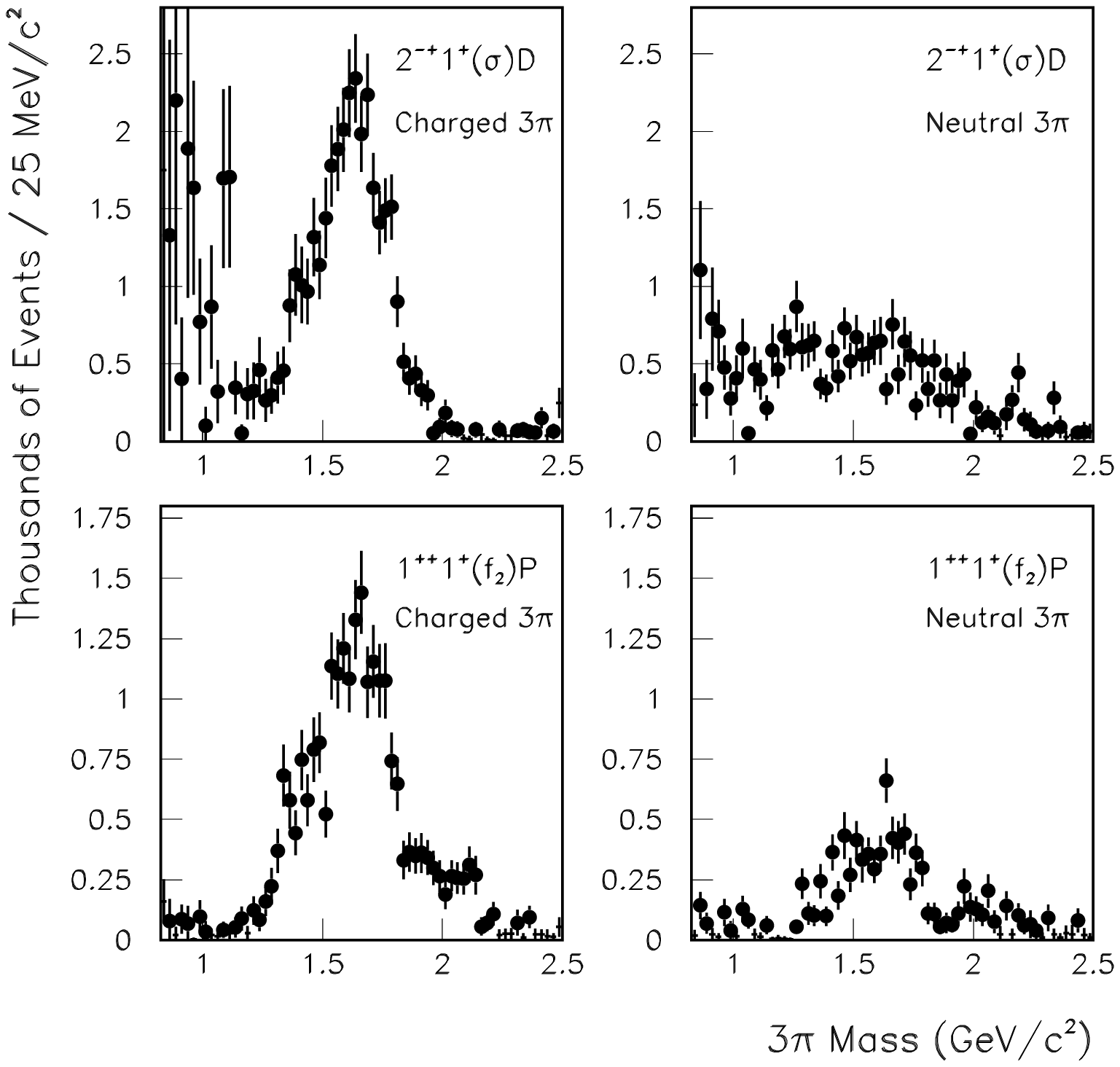}
  \caption{Intensities of the $\pitwofa$, the $\pitwofb$, the $\pitwos$, and the $\aonehigh$ waves, which were included in the high wave fit but not in the low.}
  \label{extras}
\end{figure}

A selection of partial wave intensities resulting from fits with both the high wave set and the low wave set is shown in Figure~\ref{intensities}.  The $\atwo$ (dominant mode of the $a_2(1320)$), the $\aone$ (dominant mode of the $a_1(1260)$), and the $\pitwo$ (dominant mode of the $\pi_2(1670)$) intensities show a remarkable stability to the choice of wave set.  Furthermore, the relative intensities in the two charge modes are consistent with isospin relations, which state that intensities for waves with isovector isobars ($\rho$) should be equal, and intensities for waves with isoscalar isobars ($f_2$) should appear in the charged to neutral ratio of 2:1.

The exotic $\pione$ intensity resulting from the low wave fit shows a peaking structure around 1600~$\mevcc$ (Figure~\ref{intensities}), consistent with the earlier report of the E852 Collaboration.  However, the peak is not present in the high wave fit.  Note the consistency between the two charge modes.

Phase differences between the $\onemp$, $\twomp$, and $\twopp$ waves are shown in Figure~\ref{phases}.  The phase difference between the $\twopp$ and $\twomp$ shows a rise corresponding to the $a_2(1320)$ resonance and a fall due to the $\pi_2(1670)$.  The $\onemp$ phase is found to closely track the $\twomp$ phase, indicating either the presence of a very small $\pi_1(1600)$ with nearly the same mass and width as the $\pi_2(1670)$ or indicating that the small residual background of the $\onemp$ intensity is correlated with the $\pi_2(1670)$, perhaps through leakage.

Intensities of a selection of waves included in the high wave fit but not in the low are shown in Figure~\ref{extras}.  The $\pitwofa$, the $\pitwofb$, and the $\pitwos$ waves are allowed decay modes of the $\pi_2(1670)$ and show structure consistent with the $\pi_2(1670)$ mass and width.  The $\aonehigh$ wave is a decay mode of the $a_1(1640)$~\cite{PDG}.  All of these waves peak in the 1600 to 1700~$\mevcc$ region.  When they are included in the fit, there is no peak in the $\pione$ intensity;  when they are not included in the fit, a peak appears in the $\pione$ wave (Figure~\ref{intensities}).

\section{Conclusion}

The peaking structure around 1600~$\mevcc$ in the exotic $\pione$ intensity (the purported $\pi_1(1600)$) is an artifact of the low wave fit.  When the wave set includes significant decay modes of the $\pi_2(1670)$ and the $a_1(1640)$, the peak in the exotic intensity disappears, and the small residual intensity shows no convincing evidence for resonance structure.


\begin{theacknowledgments}

This work was done in collaboration with Alex Dzierba, Adam Szczepaniak, Maciej Swat, and Scott Teige of Indiana University.

\end{theacknowledgments}



\bibliographystyle{aipproc}   

\bibliography{3pi}

\IfFileExists{\jobname.bbl}{}
 {\typeout{}
  \typeout{******************************************}
  \typeout{** Please run "bibtex \jobname" to optain}
  \typeout{** the bibliography and then re-run LaTeX}
  \typeout{** twice to fix the references!}
  \typeout{******************************************}
  \typeout{}
 }

\end{document}